\documentclass[aps,pre,reprint,noshowpacs,superscriptaddress,floatfix,letterpaper,longbibliography]{revtex4-2}
\usepackage{amsmath,amssymb,amsbsy,amsfonts,amsthm,bbm,bm,mathtools,mathrsfs,bbold}
\usepackage{color}
\usepackage{physics}
\usepackage{xfrac}
\usepackage[dvipsnames]{xcolor}
\definecolor{LapisLazuli}{RGB}{47, 102, 169}
\usepackage[hypertexnames=false,colorlinks=true,citecolor=LapisLazuli,linkcolor=LapisLazuli,urlcolor=LapisLazuli]{hyperref}
\usepackage{empheq,ragged2e}
\usepackage{pgfplots}
\usepackage{relsize}
\usepackage[inline]{enumitem}
\usepackage[normalem]{ulem}
\usepackage{comment}
\usepackage{import}

\definecolor{myred}{RGB}{214, 39, 40}
\newcommand{\jrg}[1]{\textcolor{myred}{#1}}

\usepackage{bigints}
\usepackage{soul}
\pgfplotsset{compat = newest}
\usetikzlibrary{arrows,intersections}
\usepackage{tikz-3dplot}
\usepackage{tikz}

\newcommand{\stability}{{\bm{A}}}

\newcommand{\ex}{{\bm{x}}}

\newcommand{\yu}{{\bm{u}}}
\newcommand{\brho}{{\bm{\varrho}}}
\newcommand{\bxi}{{\bm{\xi}}}

\graphicspath{{../plots/}} 

\usepackage{tcolorbox}
\usepackage{multirow}
\definecolor{new_blue}{RGB}{9, 136, 232}
\newtcolorbox{mybox}[3][]
{%
	colframe = #2!25,
	colback  = #2!10,
	coltitle = #2!20!black,  
	title    = {#3},
	#1,
}
\tcbuselibrary{breakable}

\usepackage{pgfplots}
\pgfplotsset{compat = newest}
\usetikzlibrary{arrows,intersections}
\usepackage{tikz-3dplot}
\usepackage{tikz}

\usepackage{dcolumn}

\begin{document}

\preprint{AIP/123-QED}

\def\xlist{4}
\def\ylist{4}
\preprint{AIP/123-QED} 

\title{Spectral bounds on the entropy flow rate and Lyapunov exponents\\in differentiable dynamical systems}

\author{Swetamber~Das}
\author{Jason~R.~Green}
\email{jason.green@umb.edu}
\affiliation{Department of Chemistry,\
	University of Massachusetts Boston,\
	Boston,  Massachusetts 02125, USA
}
\affiliation{Department of Physics,\
	University of Massachusetts Boston,\
	Boston, Massachusetts 02125, USA
}

\date{\today}

\begin{abstract}

Some microscopic dynamics are also macroscopically irreversible, dissipating energy and producing entropy.
For many-particle systems interacting with deterministic thermostats, the rate of thermodynamic entropy dissipated to the environment is the average rate at which phase space contracts. 
Here, we use this identity and the properties of a classical density matrix to derive upper and lower bounds on the entropy flow rate with the spectral properties of the local stability matrix. 
These bounds are an extension of more fundamental bounds on the Lyapunov exponents and phase space contraction rate of continuous-time dynamical systems.
They are maximal and minimal rates of entropy production, heat transfer, and transport coefficients set by the underlying dynamics of the system and deterministic thermostat.
Because these limits on the macroscopic dissipation derive from the density matrix and the local stability matrix, they are numerically computable from the molecular dynamics.
As an illustration, we show that these bounds are on the electrical conductivity for a system of charged particles subject to an electric field.

\end{abstract}

\maketitle

\section{Introduction}

Physical systems undergoing irreversible processes naturally produce entropy and dissipate energy~\cite{Callen1985}.
In statistical physics, there have been significant efforts to establish theoretical relationships between the entropy production and the properties of atomistic molecular dynamics.
Among these efforts is an outgrowth of dynamical systems theory~\cite{Krylov2014,dorfman1999introduction,gaspard2005chaos} that includes relationships between measures of dynamical stability and physical properties, such as energy dissipation, entropy production, and transport coefficients~\cite{Evans1990,GasNico1990,CohenRondoni1998, Ruelle1999,Qian2019,Caruso2020}.
For instance, in many-particle Hamiltonian systems exchanging energy with a Nos\'e-Hoover thermostat, the (negative) mean phase volume contraction, the Gibbs entropy production in the environment (negative entropy flow), and thermodynamic dissipation are equivalent~\cite{CohenRondoni1998,Patra2016,Ramshaw2017, Ramshaw2020}.
The Gibbs entropy rate is nonnegative for the molecular dynamics of systems deterministically thermostatted, suggesting it is a statistical analogue of the second law of thermodynamics and a bound on the irreversible thermodynamic behavior of dissipative systems~\cite{Patra2016, Ramshaw2017, Ramshaw2020, Ramshaw_2020b}.
While deterministic thermostats have been important for investigating nonequilibrium steady states~\cite{Jepps_2010}, it is less clear how to relate the microscopic dynamics to thermodynamic quantities for systems transiently out of equilibrium. 

Here, we use a classical density matrix theory~\cite{DasGreen2022} to derive upper and lower bounds on rates of the entropy flow, heat transfer, and transport coefficients in Hamiltonian systems interacting with Gaussian and Nos\'e-Hoover thermostats~\cite{Klages2007,Patra2016,Ramshaw2017}.
These bounds on dissipation, set by the deterministic dynamics, hold for systems transiently evolving through nonequilibrium states. They derive from a classical density matrix built from the tangent vectors of a dynamical system.
Tangent vectors can be simulated along with the molecular dynamics of many-particle systems; 
the full set of vectors is numerically accessible for systems with tens of thousands of particles~\cite{YangR05, GreenCGS13, CostaGreen2013}, making this density matrix theory a computationally-tractable approach to nonequilibrium statistical mechanics.
So far, the theory includes a generalization of Liouville's equation and theorem for the phase space volume of non-Hamiltonian dynamics~\cite{DasGreen2022}, as well as classical uncertainty relations and speed limits on the evolution of (observables of) differentiable dynamical systems~\cite{das2021speed,Sahbani2023,DasGreen2024}.
We add to this theory here and show that the properties of this classical density matrix make it possible to bound the phase space contraction rate, entropy production rate, and (finite-time) Lyapunov exponents.

Our approach is to first bound the Lyapunov exponents as this gives a direct route to the phase space contraction rate. 
Lyapunov exponents are well known phase space invariants, indicators of dynamical chaos, and a common tool in the analysis of complex dynamical systems~\cite{PikovskyP16, Goldhirsch1987}. 
For instance, the largest exponent characterizes the Batchelor scale of mixing in the chaotic advection of fluids~\cite{Aref2017}.
It also plays a prominent role in the turbulent dynamo action~\cite{tobias_2021}, which is responsible for generating amplified magnetic fields in astrophysical plasma, and is related to the temperature of thermalized fluids~\cite{Murga2021}. 
Finite-time Lyapunov exponents are important signatures of the bulk behavior of classical many-body systems~\cite{das_self-averaging_2017,dasExtensivityAdditivityKolmogorovSinai2017a,dascritical2019}, including those exhibiting weak, stable, or transient chaos~\cite{Szezech2005, Politi2010, Stef2010, daSilva2015}.
The largest exponent in the instantaneous Lyapunov exponent spectrum, known as the reactivity~\cite{NeuCas1997}, measures the maximum rate at which asymptotically stable systems~\cite{Lloyd2020} can transiently amplify perturbations in response to external stimuli~\cite{Lloyd1993}. 
The bounds we derive here on the entire spectrum of Lyapunov exponents hold in short and long time limits for continuous-time deterministic systems.

After summarizing the theory in Sec.~\ref{sec:dynamics_unnorm}, we put it to use in Sec.~\ref{sec:bounds_on_LE} by deriving bounds on Lyapunov exponents.
These bounds extend to the phase space volume variation rate in any deterministic dynamical system, Sec.~\ref{sec:bounds_on_dissipation}, and entropy production and heat transfer rates, Sec.~\ref{sec:bounds_entropy}.
We use models of thermostatted molecular dynamics to confirm the bounds on thermodynamic dissipation and the entropy production.
The inequalities also apply to transport coefficients, which we show by setting bounds on the electrical conductivity for a collection of charged particles in an electric field.

\section{Classical density matrix theory}\label{sec:dynamics_unnorm}

Consider a classical dynamical system with state-space variables $\{x^i\}$.
At any moment in time, these variables together mark a point $\ex(t) := [x^1(t),x^2(t),\ldots,x^n(t)]^\top$ in an $n$-dimensional state space $\mathcal{M}$ that evolves as $\dot{\ex} = \boldsymbol{F}[\ex(t)]$.
Infinitesimal perturbations to the system will also evolve under the flow of the dynamics.
Because of their analytical and computational tractability, local perturbations $\ket{\delta\ex(t)}:= [\delta x^1(t), \delta x^2(t), \ldots, \delta x^n(t)]^\top\in T\mathcal{M}$ and their linearized dynamics have become a well established means of analyzing the stability of nonlinear dynamical systems~\cite{PikovskyP16}. 
Tangent vectors, representing these perturbations to the initial condition, will stretch, contract, and rotate over time,
\begin{equation}
  \ket{\delta\dot{\ex}(t)} = \stability[\ex(t)]\ket{\delta \ex(t)},
  \label{equ:EOM_per_ket2}
\end{equation}
as they evolve with the phase point under the local stability matrix $\stability:=\stability[\ex(t)] = \grad\boldsymbol{F}$ with elements
$(\stability)^i_{j}=\partial \dot{x}^i(t)/\partial x^j(t)$.

Recently, we have taken a related approach to classical, deterministic dynamical systems, redefining the state in terms of the phase point and an associated classical density matrix~\cite{DasGreen2022}.
In this classical theory, the simplest (unnormalized) density matrix is the outer product of a single tangent vector, $\bxi(t) = \dyad{\delta \ex(t)}{\delta \ex(t)}$.
Once normalized $\brho(t)=\bxi(t)/\operatorname{Tr}(\bxi(t))$, this matrix is a projection operator with the expected properties of a density matrix: $\brho_i^2 = \brho_i$, $\Tr\brho_i = 1$, $\Tr\brho_i^2=1$, symmetric, $\brho_i \succeq 0$, i.e., $\brho_i$ is positive semidefinite. 

Impure states are also possible. For example, ``maximally mixed'' states are defined (in part) by a density matrix $\bxi^M$ composed of a complete set of linearly independent tangent vectors that span the tangent space at a phase point, $\bxi^M = \sum_{i=1}^n \bxi_i =\sum_{i=1}^n \dyad{\delta\ex_i}{\delta\ex_i}$.
These states are related to the phase space metric by a similarity transformation and have a determinant $|\bxi^M|$ representing a phase space volume element~\cite{DasGreen2022}.
For non-Hamiltonian systems, the phase space volume contraction rate $\Lambda$ determines the time evolution of this state, which is a generalization of Liouville's theorem and equation~\cite{DasGreen2022}.

Known observables, including Lyapunov exponents and the phase space contraction rate, are expectation values over classical density matrices.
To make this point and the similarity with the quantum density matrix more apparent, we can define the classical density matrix with unit tangent vectors.

First, we define the $i^{\text {th}}$ tangent vector as $\ket{\delta \yu_i} = \ket{\delta \ex_i}/\|\delta \ex_i\|$, where $\|.\|$ is the $\ell_2$-norm and we treat the phase space variables as dimensionless. 
The equation of motion for this unit vector,
\begin{align}
\label{equ:EOM_unit_per}
  \frac{d}{dt}\ket{\delta \yu_i} &= \stability\,\ket{\delta \yu_i} - r\ket{\delta \yu_i},
\end{align}
contains a source/sink term with the instantaneous rate:
\begin{align} \label{eq:ILE}
	r_i:=r_i(t) = \bra{\delta \yu_i}\stability_+\ket{\delta \yu_i}= d_t\ln\|\delta \ex_i\|.
\end{align}
This instantaneous Lyapunov exponent (or local stretching rate) for the linearized dynamics is coordinate dependent~\cite{EckYao1993, WalKla2012} and depends on the symmetric part of the stability matrix, $\stability_+ = (\stability+\stability^\top)/2$.
At every phase point of an $n$-dimensional phase space, there are $n$ instantaneous exponents.
Each element of this spectrum can be time averaged to define the corresponding finite-time Lyapunov exponent,
\begin{align}\label{eq:FTLE_def}
  \lambda_i(t) := \lambda_i(t,t_0) = |t-t_0|^{-1}\int_{t_0}^tr_i(t)\,dt.
\end{align}
In the long-time limit, these become the Lyapunov exponents, $\lambda=\lim_{t\to\infty}\lambda_i(t)$, which are independent of time and the choice of initial condition for a given regular or chaotic trajectory~\cite{Ott_2002}.

With this context, we can see the alternative representation of Lyapunov exponents in this classical density matrix theory: Each Lyapunov exponent is the expectation value~\cite{DasGreen2022} computed with respect to a density matrix constructed from a unit tangent vector, $\brho_i(t) = \dyad{\delta \yu_i(t)}{\delta \yu_i(t)}$.
For this state $\brho_i$, the instantaneous Lyapunov exponent defined in Eq.~\ref{eq:ILE} is the tangent space average of $\stability_+$: $r_i = \Tr(\stability_+\brho_i)= \langle\stability_+\rangle_{\brho_i}$. This observation suggests that other observables might be expressed as expectation values over a classical density matrix at a given point in time, $\langle \boldsymbol{O}\rangle_{\brho_i}= \Tr(\boldsymbol{O}\boldsymbol{\brho}_i)$. 

Second, the phase space variation (or contraction) rate is also an expectation value over a maximally mixed classical density matrix.
The density matrix $\brho_i$ for each tangent vector evolves in time according to
\begin{equation}\label{equ:EOM_rho}
\frac{d\brho_i}{dt} = \{\stability_+,\brho_i\} + [\stability_-,\brho_i] - 2\langle\stability_+\rangle_{\brho_i}\,\brho_i,
\end{equation} 
a classical analogue of the von-Neumann equation with an anti-commutator $\{\cdot\}$ and a commutator $[\cdot]$.
The last term on the right is a source/sink term with the associated instantaneous Lyapunov exponent.
The phase space variation rate is the sum of these exponents $\Lambda(t) = \sum_{i=1}^n r_i(t) = \sum_{i=1}^n\langle\stability_+\rangle_{\brho_i}$.

Another representation of the phase space variation rate is as an expectation value over the maximally mixed state. 
Take a complete set of $n$ linearly independent unit tangent vectors.
If these vectors span a phase space volume around a phase point $\brho^M(t) = n^{-1} \sum_{i=1}^n\brho_i(t)$, the density matrix $\brho^M$ represents a maximally mixed state.
As in quantum dynamics, a complete set of pure state states contribute with equal weights, $n^{-1}$, ensuring $\Tr \brho^M= 1$.
The local phase space variation (expansion or contraction) rate is the expectation value of $\stability_+$ over a maximally mixed state $\brho^M$: $\Lambda = n\langle\stability_+\rangle_{\brho^M}=n\Tr(\stability_+\brho^M) = \sum_{i=1}^n\Tr(\stability_+\brho_i) = \sum_{i=1}^n\langle\stability_+\rangle_{\brho_i}$.
So while the instantaneous Lyapunov exponents $r_i$ are expectation values over pure states $\brho_i$, the phase space contraction rate is an expectation value over a maximally mixed state $\brho^M$.

Given two well known observables are expectations values over a classical density matrix suggests a closer look at how these observables are related to the dynamical stability encoded in the stability matrix $\stability$. 
With only the ingredients of the density matrix approach above, we can show that the spectral properties of the density matrix $\brho_i$ and the stability matrix $\stability_+$ set upper and lower bounds on any instantaneous exponent in the Lyapunov spectrum.

\section{Upper and lower bounds on local and finite-time Lyapunov exponents}\label{sec:bounds_on_LE}

\begin{figure*}[t]
	\centering
	\hspace*{-0.7cm}\includegraphics[width=0.8\textwidth]{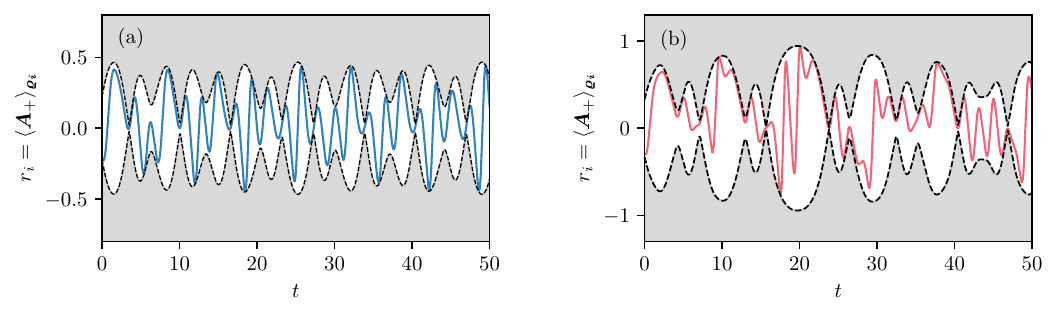}
	\caption{\label{fig:ILE_bounds} For the H\'enon-Heiles system, the minimum and maximum eigenvalues of symmetric part of the stability matrix, $\stability_+$, are $\pm\sqrt{x^2+y^2}$. These eigenvalues set the upper and lower bounds (Eq.~\ref{eq:ILE_bound}) on the instantaneous Lyapunov exponent $r$ for (a) a regular orbit (blue) with $E = 1/12$ and (b) a chaotic orbit (red) with $E = 1/6$. The shaded region is inaccessible to $r$ at any time.}
\end{figure*}

Finite-time Lyapunov exponents are time averages of the instantaneous or local Lyapunov exponent $r_i$, both of which depend on the choice of coordinates and typically fluctuate along a trajectory~\cite{WalKla2012}.
Nevertheless, bounds on these observables are possible (in any basis) expressing the observables as the trace.
For instance, we can use $r_i = \langle \stability_+\rangle_{\brho_i}$ to derive upper and lower bounds upon recognizing $\Tr(\stability_+\brho_i) = \bra{\delta \yu_i}\stability_+\ket{\delta \yu_i}$ as the Rayleigh quotient of $\ket{\delta \yu_i}$ with respect to the symmetric matrix $\stability_+$.
Given the diagonalizability of $\stability_+$ and the positive-semidefiniteness of $\brho_i$, the minimum, $\sigma_\text{min}^{\stability_+}$ and maximum eigenvalues $\sigma_\text{max}^{\stability_+}$ of $\stability_+$ set bounds on $r_i=\langle \stability_+\rangle_{\brho_i}$,
\begin{align}\label{eq:ILE_bound0}
   \sigma_\text{min}^{\stability_+}\leq  \frac{\Tr(\stability_+\brho_i)}{\Tr(\brho_i)} \leq\sigma_\text{max}^{\stability_+}.
\end{align}
according the min-max theorem~\cite{horn1990matrix}.
The upper (lower) bound saturates when $\brho_i$ is composed of the eigenvector of $\stability_+$ which corresponds to its maximum (minimum) eigenvalue.
In the numerator of Eq.~\ref{eq:ILE_bound0}, both $\stability_+$ and $\brho_i$ are symmetric but only $\brho_i$ is generally positive semidefinite.
These inequalities have been rigorously proven~\cite{Wang1986} for any two real matrices $\boldsymbol{X}$ and $\boldsymbol{Y}$ of order $n \times n$, provided $\boldsymbol{X}$ is symmetric and $\boldsymbol{Y}$ is positive semidefinite, $\sigma_\text{min}^{\boldsymbol{X}} \Tr(\boldsymbol{Y}) \leq \Tr(\boldsymbol{X}\boldsymbol{Y}) \leq \sigma_\text{max}^{\boldsymbol{X}} \Tr(\boldsymbol{Y})$.
In this general case, the minimum $\sigma_\text{min}^{\boldsymbol{X}}$ and maximum $\sigma_\text{max}^{\boldsymbol{X}}$ eigenvalues of $\boldsymbol{X}$ set the bounds.

Because dynamics preserve the trace $\Tr\brho_i = 1$ at all times, the inequalities simplify to upper and lower bounds on the instantaneous Lyapunov exponent,
\begin{align}\label{eq:ILE_bound}
   \sigma_\text{min}^{\stability_+}(t)\leq  r_i(t) \leq\sigma_\text{max}^{\stability_+}(t),
\end{align}
which, again, is $r_i = \Tr(\stability_+\brho_i)$.
To emphasize, the unit trace of $\brho_i$ for all times is the key in setting these bounds with only the extremal eigenvalue of $\stability_+$.
The instantaneous Lyapunov exponents are time dependent and the bounds hold at all times.
One can also derive this particular version of the upper bound directly from the time derivative of the $\ell^2$ norm of $\ket{\delta \ex}$ and Rayleigh's principle~\cite{NeuCas1997}.

The upper (or lower) bound in Eq.~\ref{eq:ILE_bound} saturates when the chosen state $\brho_i$ is along the eigenvector of $\stability_+$ with largest (or smallest) eigenvalue, respectively.
In other words, the extremal eigenvalues of $\stability_+$ represent the maximum possible rates of local stretching and contraction for a tangent vector at any given point along a trajectory.
In a chaotic system, as a tangent vector evolves in time, it ultimately converges onto the eigenvector of $\stability$ which has the largest real eigenvalue.
However, the maximum local stretching (or contraction) occurs along the eigenvectors of $\stability_+$ with the largest (or smallest) eigenvalues, respectively.

Time averaging along a trajectory, these inequalities in Eq.~\ref{eq:ILE_bound} extend directly to any finite-time Lyapunov exponent, $\lambda_i(t)$ using their definition from Eq.~\ref{eq:FTLE_def}:
\begin{align}\label{eq:LE_bounds}
\overline{\sigma}_\text{min}^{\stability_+}(t) \leq \lambda_i(t) \leq \overline{\sigma}_\text{max}^{\stability_+}(t).
\end{align}
The overbar here, and throughout, indicates the time average $\overline{O} = t^{-1}\int_{0}^{t}dt'O$.
While $r_i$ depends on the density matrix $\brho_i$, their extremal values are always the minimum and maximum eigenvalues of $\stability_+$.
These spectral limits, both those in Eqs.~\ref{eq:ILE_bound} and \ref{eq:LE_bounds}, are time-dependent bounds on the evolution of local and finite time perturbations in a deterministic system.

Both sets of bounds, Eqs.~\ref{eq:ILE_bound} and~\ref{eq:LE_bounds}, are numerically, and in some cases, analytically, computable for differentiable deterministic systems. 
As numerical confirmation, we simulated the H\'enon-Heiles system, which is a minimal model for the planar motion of stars around a galactic center~\cite{Henon1964}.
Using the potential $V(x,y)= \frac{1}{2}(x^2+y^2 + 2x^2 y - \frac{2}{3}y^3)$, the equations of motion are
\begin{align}
\dot{x} &= p_x,   \quad \quad \quad \quad \dot{y} = p_y,\nonumber\\
\dot{p_x} &= -x - 2xy,  \quad \dot{p_y} = -y - x^2 + y^2.
\end{align}
The maximum and minimum eigenvalues of $\stability_+$ computed from these equations are $\pm\sqrt{x^2+y^2}$.
Figure~\ref{fig:ILE_bounds} shows these bounds for representative regular and chaotic trajectories at energies $E = 1/12$ and $1/6$, respectively.
To compute the instantaneous exponents, we first generate a phase space trajectory and initiate a normalized tangent vector with random elements at any point on the trajectory.
We use this random tangent vector to construct the density matrix $\brho_i$ and evolve it according to Eq.~\ref{equ:EOM_rho}.
We then compute $\Tr (\stability_+\brho_i)$ at each point on the trajectory.
 To show the  bounds, we compute the extremal eigenvalues of $\stability_+$ at those points.
The inequalities in Eq.~\ref{eq:ILE_bound} hold for all four exponents, but we only show the largest one in Fig.~\ref{fig:ILE_bounds}.
These initial analysis of the H\'enon-Heiles dynamics prompted a closer look at Hamiltonian dynamics that gives an interpretation of the upper and lower limits.

For Hamiltonian dynamics, the inequalities bracketing the instantaneous Lyapunov exponents in Eq.~\ref{eq:ILE_bound} are directly related to the curvature of phase space.
Given a dynamical system with a Hamiltonian of the form $H(q,p) = T(p) + V(q)$, these exponents are the maximum and minimum eigenvalues of the symmetric part of the stability matrix, $\pm \frac{1}{2}(\partial_p^2 T-\partial_q^2 V)$.
In the case of \textit{natural} Hamiltonian systems, whose kinetic energy is quadratic in the velocities, the bounds are related to the potential energy and mass.
This connection is explicit in the geometric theory of chaotic dynamics that identifies phase space space trajectories as geodesics in the configuration space of the system~\cite{Casetti2000, Horwitz2007}.
In this theory, the stability of trajectories is linked with the stability of geodesics and determined by the curvature of a Riemannian manifold.
The average fluctuations of the curvature gives an estimate of the largest Lyapunov exponent, with the curvature of the manifold being the curvature of the potential $\partial^2_q V$.
Here, this curvature appears in the upper (or lower) bound on the local Lyapunov exponent along a trajectory, making these bounds a consequence of the local fluctuations in the curvature of the configuration space manifold. 
Moreover, if the system has a single particle of mass $m$, then $\partial^2_p T \propto m^{-1}$.
Thus, both curvature of the potential and mass (in scaled units) directly contribute to the limiting rates in these systems.

The extremal eigenvalues of $\stability_+$ define a set of time scales (given by the inverse of those eigenvalues) for the evolution of dynamics systems over finite times.
These time scales could be important for transient phenomena, for example, to characterize non-normality-induced temporal perturbation growth and to characterize transient chaos~\cite{Tel2015}, stable chaos~\cite{Politi2010}, collective chaos~\cite{Politi2017}.
To analyze the use of these bounds for transient processes, we consider a dynamically unstable model: the inverted harmonic potential $H(q,\,p) = \frac{1}{2}(p^2/m - \kappa q^2)$ which in rescaled phase space variables takes the form $H(x,y) = \sqrt{\kappa/m}xy$~\footnote{The Hamiltonian for the inverted harmonic oscillator: $H(q,p) = \frac{1}{2}(p^2/m - \kappa q^2) =  \frac{1}{2}(p^2/m - m\omega^2 q^2) $. Using a canonical change of coordinates $q = 1/\sqrt{2m\omega}(x-y)$ and $p = \sqrt{m\omega/2}(x+y)$, the Hamiltonian $H(q,\,p) = \frac{1}{2}(p^2/m - \kappa q^2)$ is transformed to $H(x,y) = \sqrt{\kappa/m} x y$. The Lyapunov exponent is then $\sqrt{\kappa/m}$}.

For this system, the Lyapunov exponents are also closely related to the relaxation and decay rates through Ruelle-Pollicott (RP) resonances~\cite{Pollicott1985,Ruelle1987,GasDorf1995}.
All trajectories are unstable, with the parameter $\kappa/m$ setting the positive Lyapunov exponent corresponding to the decay lifetime of the correlations, $\lambda:=\sqrt{\kappa/m}= \tau_D^{-1}$~\cite{Gas2006}.

With this setup, we can apply the density matrix theory. 
Defining the classical density matrix as 
\begin{align}
\brho = \frac{1}{\delta x^2 + \delta y^2}
\begin{pmatrix}
\delta x^2 &	\delta x\delta y\\ 
\delta x\delta y & \delta y^2
\end{pmatrix}, 
\end{align}
the instantaneous Lyapunov exponent is
\begin{align}
\label{eq:11}
r = \langle\stability_+\rangle_\brho = \lambda\left(\frac{\delta x^2 \,- \delta y^2}{\delta x^2\, +\delta y^2}\right).
\end{align}
The resonances for the inverted harmonic potential are integer multiples of $\lambda$~\cite{Gas2006}: $s_{lg} = -(l + g +1)\lambda$, with integers $l,g = 0,1,2,\ldots$.
The leading resonance $s_{00}$ with $l=g=0$ is $-\lambda$.
The trajectories of this system exhibit transient dynamics characterized by correlations that decays asymptotically at a rate $s_{00}$.
For individual trajectories, however, local stretching rates fluctuate in finite time within the bounds set by $\lambda=\sqrt{\kappa/m}$:
\begin{align}
  -\sqrt{\frac{\kappa}{m}} = s_{00} \leq r \leq \sqrt{\frac{\kappa}{m}} = -s_{00},
\end{align}
where $s_{00}$ is time independent.
The bounds on the instantaneous exponent $r$ hold during transient dynamics. The value of $r$ could be negative or positive depending on the sign of $(\delta x^2 - \delta y^2)$ in Eq.~\ref{eq:11}. 
Thus, for natural Hamiltonian systems, the bounds on the instantaneous Lyapunov exponents and RPs are directly set by mass and the second derivative of the potential $-\partial^2_qV = \kappa$.

The applicability of the bounds in Eq.~\ref{eq:ILE_bound} to transient phenomena suggest they may be useful in understanding nonequilibrium processes dissipating energy and producing entropy.
To put bounds on the entropy production, we first extend our results to the rate of phase space contraction characteristic of dissipative systems~\cite{Ott_2002}.

\section{Upper and lower bounds on the phase space variation rate}\label{sec:bounds_on_dissipation}

The bounds on finite-time Lyapunov exponents are a theoretical basis for limits on the rate of entropy production and flow.
To set these limits, we first extend the inequalities in Eq.~\ref{eq:ILE_bound} to the rate of phase space variation.
In dissipative systems, the phase space volume typically contracts with time as a result of energy being expelled to the surroundings~\cite{Argyris2015}.
For instance, take the damped harmonic oscillator with unit mass and unit frequency and the energy function $E(q,p, \gamma)=\frac{1}{2}(p^2 + q^2) - \gamma p$.
The rate of total energy dissipation is $\gamma p^2$.
The damping coefficient $\gamma$ is proportional to the rate at which local phase space volume $\delta q\delta p$ is compressed and a measure of local energy dissipation along a phase space trajectory.

To place upper and lower bounds on the phase volume contraction rate, we start with the sum of all the instantaneous Lyapunov exponents, which determines this rate at a given point on the trajectory, $\Lambda = \sum_{i=1}^n r_i$~\cite{Eckmann1985Ergodic, PoschHoover2006}.
Equation~\ref{eq:ILE_bound} provides bounds on each of these individual exponents.
Summing those inequalities gives basis-independent bounds on $\Lambda$,
\begin{align}\label{eq:bound_contraction_rate}
n\sigma_\text{min}^{\stability_+}(t)\leq \Lambda(t)\leq n\sigma_\text{max}^{\stability_+}(t),
\end{align}
anywhere in the $n$-dimensional phase space.
As before, time averaging
\begin{align}\label{eq:avg_dissipation}
n\overline{\sigma}_\text{min}^{\stability_+}\leq \bar\Lambda \leq n\overline{\sigma}_\text{max}^{\stability_+}
\end{align}
leads to upper and lower bounds on the time-average phase space contraction rate, $\bar{\Lambda}$.
Here, dynamical quantities $\sigma^{\stability}$ bound a physically significant parameter.

To see this physical significance, consider the unforced van der Pol oscillator~\cite{van-der-pol},
\begin{align}
\dot x = y, \quad \dot y = -x - \mu(x^2-1)y,
\end{align}
which a non-conservative system exhibiting self-sustained oscillations.
The nonlinear damping strength $\mu > 0$ controls the qualitative features of the stable limit cycle.
What motivates this example here is that the local energy dissipation along the limit cycle is the local phase space contraction rate: $\Lambda = -\mu(x^2-1)$.
This contraction rate also affects the rate of change of energy $d_tE= x \dot x + y\dot y = \Lambda y^2$. 
It is also the rate at which an infinitesimal volume element (collection of phase space points) collapses in time
$d_t\ln (\delta p \delta q)$ because the local energy dissipation (local to a single trajectory) is $d_t(\delta p \delta q) = \Lambda  (\delta p \delta q) $. 
This local energy dissipation $d_t(\delta p \delta q)$ has dimensions of energy -- that is, the local phase space volume contraction rate sets the local energy dissipation. The local energy dissipation $\epsilon := d_t(\delta q\delta p) = d_t\delta V=  (\nabla\cdot\dot{\ex}) \delta V$ suggests a kind of ``energy density'' $\epsilon/\delta V = \nabla\cdot\dot{\ex}$.
Our bounds here are on $\epsilon/\delta V$.

\begin{figure}[t]
	\centering
	\hspace*{-0.75cm}\includegraphics[width=0.45\textwidth]{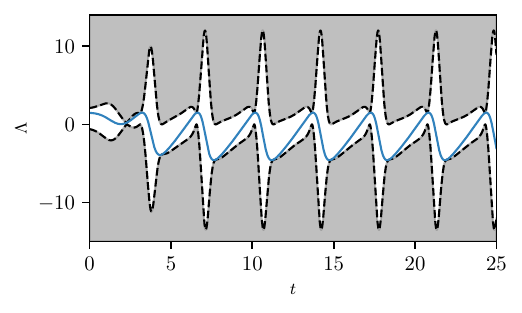}
	\caption{The van der Pol oscillator: Upper and lower bounds (dashed black lines) $(\Lambda \pm \sqrt{\Lambda^2 + 4\mu^2x^2y^2})/2$ on $\Lambda/n$ (blue line) for $\mu = 1.5$. The shaded regions are inaccessible to $\Lambda/n$. The state space is two dimensional, $n = 2$.}
	\label{fig:vDP-bounds}
\end{figure}

\begin{figure}[t]
	\centering
	\hspace*{-0.4cm}\includegraphics[width=0.45\textwidth]{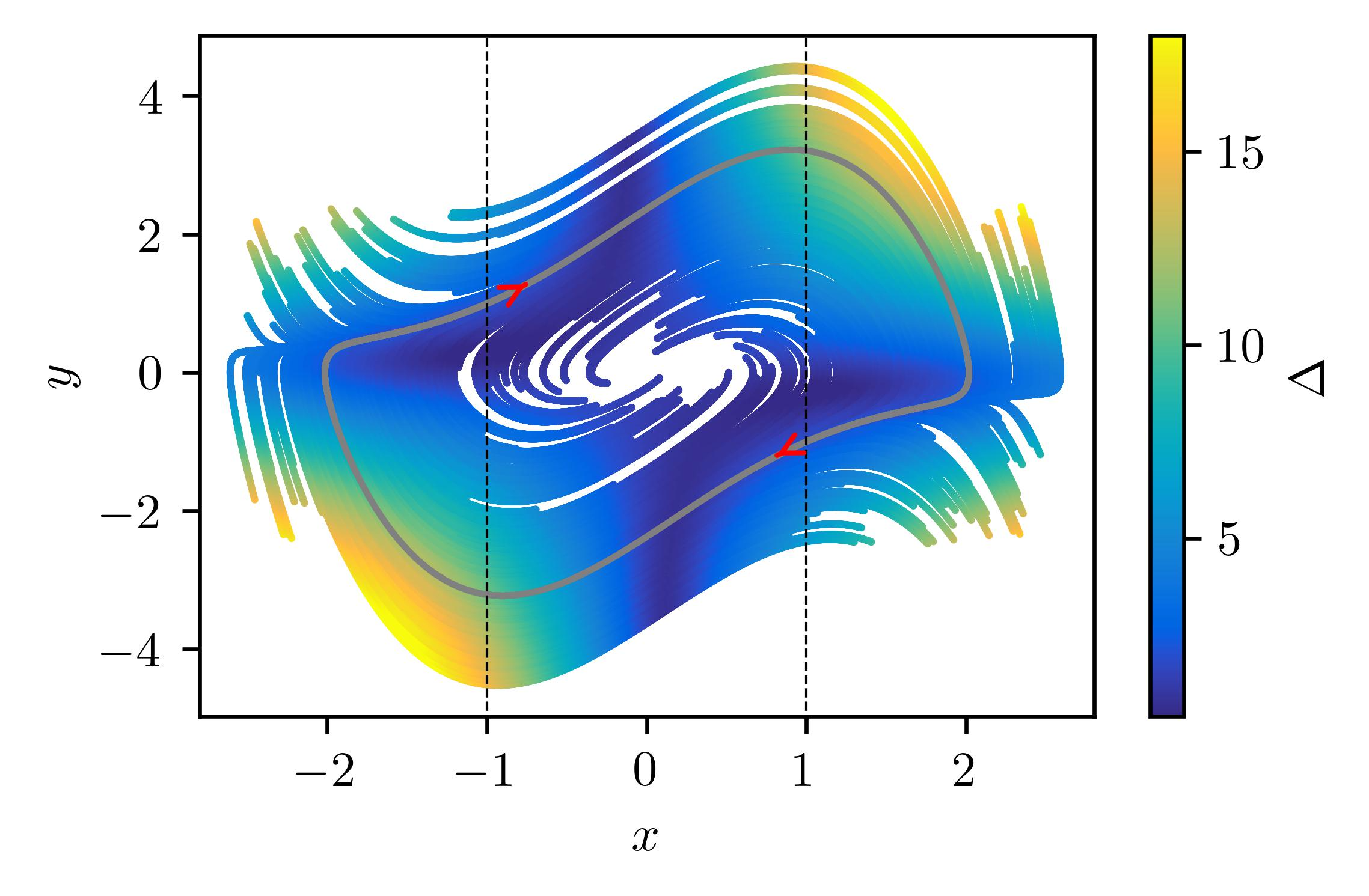}
	\caption{Phase space of the van der Pol oscillator at $\mu = 1.5$: a set of 500 different trajectories color coded according to the deviation of the local dissipation rate $\Lambda$ from lower (upper) bound given by  the smallest (largest) eigenvalue of the associated stability matrix. These deviations at a given time are the same, i.e., $\Delta = (\sqrt{\Lambda^2 + 4\mu^2x^2y^2})/2$. For $-1\leq x\leq 1$, $\Lambda > 0$. All of these trajectories approach the limit cycle (gray).}
	\label{fig:vDP-dev-bounds}
\end{figure}

The van der Pol oscillator experiences a continuous gain/loss of energy at low/high amplitude. Its steady state has sustained periodic oscillations and is net dissipating.
At high amplitudes ($|x| > 1$), the van der Pol oscillator is damped, $\Lambda < 0$.
At low amplitudes ($|x| < 1$), it accumulates energy $\Lambda > 0$.
During these oscillations, the phase space variation rate also oscillates, which makes both the upper and lower bounds here nontrivial.
According to Eq.~\ref{eq:bound_contraction_rate}, the upper and lower bounds on energy dissipation (generation) rate are $(\Lambda \pm \sqrt{\Lambda^2 + 4\mu^2x^2y^2})/2 $.
Figure~\ref{fig:vDP-bounds} shows the bounds on $\Lambda$ for a density matrix $\brho$ built from an arbitrary unit tangent vector.
The contraction rate $\Lambda/2$ scaled by one half deviates from the upper and lower bounds by the same amount: $\Delta = (\sqrt{\Lambda^2 + 4\mu^2x^2y^2})/2$.
A set of $500$ trajectories is shown in Fig.~\ref{fig:vDP-dev-bounds} color coded according to the magnitude of $\Delta$.
Deviations $\Delta$ become smaller at low amplitudes ($|x|<1$, $\Lambda> 0$), where energy is gained, as compared to high amplitudes ($|x|>1$, $\Lambda< 0$), where energy is lost.
In Fig.~\ref{fig:vDP-dev-bounds}, energy is gained when trajectories visit the phase space region $-1<x<1$ marked \jrg{by} the vertical dashed lines, and energy is lost for trajectories outside of this region.


Overall, the bounds on $\Lambda$ hold for the phase space contraction rate at any moment in time along a trajectory, regardless of whether the system is in a steady-state or not.
Provided the dynamics are differentiable, the system could also be open or closed, evolving passively or driven actively.

The bounds on the Lyapunov exponents also serve as a building block for bounds on other quantities. 
For example, instead summing Eq.~\ref{eq:ILE_bound} over the positive instantaneous Lyapunov exponents gives bounds on the \textit{finite-time} Kolmogorov-Sinai (KS) entropy $h_{\scriptscriptstyle\text{KS}}$ rate. According to Pesin's theorem~\cite{Ott_2002}, its long-time limit is the sum of positive Lyapunov exponents for isolated and closed systems.
For a $2n$-dimensional Hamiltonian system with $n$ positive exponents,
\begin{align}\label{eq:KS_bounds}
\frac{h_{\scriptscriptstyle \text{KS}}}{n} \leq {\sigma}_\text{max}^{\stability_+}.
\end{align}
We omit the lower bound, which is trivial in this case because the smallest eigenvalue $\sigma_\text{min}^{\stability_+} < 0$.
For open dynamical systems, the sum of positive exponents or local dispersion rate provides useful connections with transport and reaction-rate coefficients through the escape-rate formalism~\cite{DorfGas1995,GasDorf1995,DorfmanBeijeren1997,Gas2006,Gas2015}.

Next, we use the result in Eq.~\ref{eq:avg_dissipation} to place bounds on Gibbs entropy rate.

\section{Upper and lower bounds on the Gibbs entropy rate}\label{sec:bounds_entropy}

Gibbs' entropy is a statistical quantity characterizing the dynamics of macroscopic physical systems~\cite{Goldstein2019}.
For a statistical phase space density $\rho(\ex,t)$, the Gibbs entropy is~\cite{And1985}
\begin{align}
  \textit{S}(t) = -\int_{\mathcal{M}} d\ex \rho(\ex,t)\ln \rho(\ex,t).
\end{align}
The probability density $\rho(\ex,t)$ is the solution of the Liouville's equation $\partial_t\rho + \nabla\cdot(\dot{\boldsymbol{x}}\rho)=0$.
Its time derivative, the Gibbs entropy rate $d_t \textit{S}$, is equal to the mean phase space contraction rate $\langle\Lambda(t)\rangle$~\cite{Daems1999,Gilbert1999,BeijerenDorfman2000}: $d_t \textit{S}(t) = \langle\Lambda(t)\rangle = \int_{\mathcal{M}} d\ex\rho(\ex,t)\Lambda$.
Here, $\langle \cdot\rangle$ indicates the average over the statistical density.
As shown by Andrey~\cite{And1985}, this identity between the Gibbs entropy rate and phase space volume contraction rate is valid for any time $t$.

Like $\Lambda(t)$, the Gibbs entropy rate $d_t\textit{S}(t)$ does not have a definite sign.
However, in dissipative systems, the time-averaged rate of phase volume contraction is non-positive in the asymptotic time limit, $\overline{\langle\Lambda\rangle}^\infty = \lim_{t\rightarrow \infty} t^{-1}\langle \Lambda \rangle \leq 0$. Systems out of equilibrium settle to a steady state at long times. 
To ensure the positivity of the average Gibbs entropy (production) rate for nonequilibrium steady states, Ruelle~\cite{Ruelle1996, Ruelle1997} proposed a sign convention
\begin{align}
\overline{d_t\textit{S}}^\infty = -\overline{\langle \Lambda\rangle}^\infty\geq 0,
\end{align}
where $\overline{d_t\textit{S}}^\infty$ indicates the average Gibbs entropy rate in the asymptotic time limit.
This convention suggests assigning the negative of $\langle\Lambda(t)\rangle$ to be the rate of entropy flowing to the environment.

Adopting this convention and using the relationship between the entropy and phase space contraction rate, we can establish bounds on $d_t\textit{S}(t)$ by averaging Eq.~\ref{eq:bound_contraction_rate} over the entire phase space with respect to the statistical density $\rho$:
\begin{align}\label{eq:Entropy_prod_prod_space_avg}
-\langle\sigma_\text{max}^{\stability_+}(t)\rangle \leq \frac{d_t\textit{S}(t)}{n} \leq -\langle\sigma_\text{min}^{\stability_+}(t)\rangle.
\end{align}
Because $d_t\textit{S} = - \langle\Lambda\rangle$, we have switched signs and reversed the direction of the inequalities.
Again, these inequalities hold at any arbitrary time $t$ and, so, they can be time averaged to produce finite time or asymptotic bounds. For finite-time averages, they are:
\begin{align}\label{eq:time-avg-bounds}
-\overline{\langle\sigma_\text{max}^{\stability}\rangle} \leq \frac{\overline{d_t\textit{S}}}{n} \leq -\overline{\langle\sigma_\text{min}^{\stability_+}\rangle}.
\end{align}
where the overbar indicates a finite time average: $\overline{O} = t^{-1}\int_{0}^{t}dt'O$. The asymptotic time average can be obtained by taking the limit $t \to \infty$ for each term in the inequalities. 

Unlike the second law of thermodynamics, which states that in isolated systems, the total entropy change is nonnegative, $\overline{d_t\textit{S}}^\infty \geq 0$, here we find, in the asymptotic time limit, a potentially finite lower bound on the entropy rate, $\overline{d_t\textit{S}}^\infty \geq -n\overline{\langle \sigma_{\max}^{\stability_+}\rangle}^\infty$.
If the dynamics are chaotic, then this lower bound is trivial. 
It will be nontrivial for dynamics with a largest Lyapunov exponent that is negative.
For example, neural networks and cellular automata are possible dynamics~\cite{Politi2010} where the bound could be applied if it extends to spiky dynamics and discrete maps. 
There is also an upper bound on the entropy flow rate:
the ensemble average of the smallest (negative) eigenvalue of $\stability_+$ is the maximal rate at which entropy can be produced, $\overline{d_tS} \leq - n\overline{\langle\sigma_{\min}^{\stability_+}\rangle}^\infty$.
This bound holds for nonequilibrium steady states and has no thermodynamic counterpart.
Both bounds hold for non-Hamiltonian systems and apply to an ensemble of systems, even when the underlying dynamics are driven or transient.

For ergodic systems, one can use the equality of phase space averages and sufficiently long time averages.
In this case, we can use time averages to compute the bounds.
Thus, both upper and lower limits are determined by the underlying microscopic dynamics and dimensionality of the phase space.
The Gibbs entropy is extensive in system size, so the factor of $n^{-1}$ ensures the quantities here are intensive.
To illustrate these bounds, we analyze the entropy rate for an ergodic thermostat and the electrical conductivity of the driven Lorentz gas in two dimensions.

\begin{figure}[t!]
	\hspace{-0.5cm}\includegraphics[width=1\columnwidth]{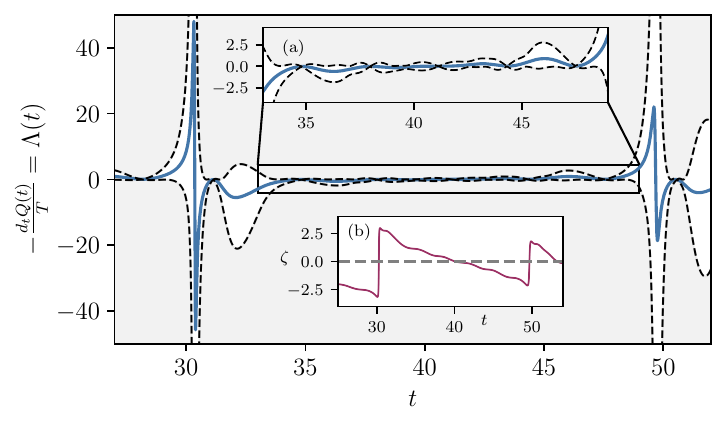}
	\caption{\label{fig:0532-lambda} The rate of heat exchange $d_t Q/T$ of a 0532 model is bounded by the extremal eigenvalues of the symmetric part of the stability matrix, Eq.~\ref{eq:thermostat_0532}, at all times during a transient phase trajectory. Parameters here are $\epsilon = 0.25$ and $T = 1$. We show the time interval $27\leq t \leq 52$ for illustration. The shaded region cannot be accessed by $d_t\textit{Q}$. \textit{Inset}. --- (a) Magnified interval $33\leq t\ \leq 49$ to show the sharpness of the bounds. (b) The $\zeta$ values in the time interval $27\leq t \leq 52$ of the same trajectory. The abrupt switching between $\zeta < 0$ to $\zeta > 0$ halves generates the two large spikes in $\Lambda$ in the main plot. The phase space variables $q$ and $p$ exhibit oscillatory behavior (not shown).}
\end{figure}

\subsection{Thermostatted, heat-conducting oscillator}
\label{sec:thermostat}

Deterministic time-reversible thermostats have direct links between phase volume contraction rate, thermodynamic dissipation, entropy production rate, and transport coefficients~\cite{Klages2007}.
Thermostats for Hamiltonian systems introduce fictitious forces to mimic the thermodynamics forces required to sustain nonequilibrium steady states~\cite{Jepps_2010, Patra2015,HariPatra2021}, usually by maintaining either the kinetic energy (isokinetic) or the total energy of the system (isoenergetic). 
In systems subject to Gaussian and Nos\'e-Hoover thermostats, the Gibbs entorpy rate, the mean phase space volume contraction rate, and the thermodynamic entropy production are identical~\cite{Ramshaw2020, Patra2016}:
\begin{align}\label{eq:diss-entropy-heat-avg}
\overline{d_t\textit{S}}^\infty = -\overline{\Lambda}^\infty= \overline{\frac{d_tQ}{T}}^\infty.
\end{align}
A phase space average of $\Lambda$ is unnecessary here because when the system is in a steady state, all phase space trajectories are on the same attractor.
For isokinetic thermostats, this identity is valid for any number of interacting particles in the system~\cite{CohenRondoni1998}.
And, in the case of a \textit{thermostatted} classical ideal gas, the relation holds for any number of gas particles because there are no interactions (isokinetic and isoenergetic thermostatting are equivalent in this case). 

Because the time averaged rates of thermodynamic entropy and Gibbs entropy are equal in these thermosttated systems, we have bounds on the heat transfer rate $\overline{d_tQ}^\infty$ that follow from from Eq.~\ref{eq:time-avg-bounds}:
\begin{align}\label{eq:heat-transfer}
-nT\overline{\sigma_\text{max}^{\stability_+}}^\infty \leq \overline{d_t \textit{Q}}^\infty\leq -nT\overline{\sigma_\text{min}^{\stability_+}}^\infty.
\end{align}
To illustrate these results, we first consider a one-dimensional oscillator with coordinate $q$ and momentum $p$ interacting with a thermostat~\cite{Hoover2016a, Hoover2016b, Patra2016}.
The oscillator experiences a linear friction force $-\zeta p$ at temperature $T$, giving equations of motion,
\begin{equation}\label{eq:thermostat_0532}
\begin{aligned}
\dot{q} &= p, \quad \dot{p} = -q - \zeta\left(\alpha p + \frac{\beta p^3}{T}\right),\\
\dot{\zeta} &= \alpha\left(\frac{p^2}{T} - 1 \right) + \beta \left(\frac{p^4}{T^2} -  \frac{3p^2}{T}\right).
\end{aligned}
\end{equation}
The variable $\zeta$ stabilizes the kinetic energy of the system.
With parameters $\alpha=0.05$ and $\beta=0.32$, this is the ``0532 model''~\cite{Hoovers2020, Hoover2016b}.
At thermal equilibrium, its stationary distribution is to the product of Gibbs canonical ensemble and a Gaussian distribution for $\zeta$: $\rho(q,p,\zeta)\propto e^{-q^2/2T}e^{-p^2/2T}e^{-\zeta^2/2T}$ in units with $k_B=1$~\cite{Hoovers2020}.

When $\epsilon\neq 0$, a temperature gradient drives the system out of equilibrium. 
The temperature profile $T = 1 + \epsilon\tanh(q)$ with $0\leq\epsilon< 1$, so that the system is at equilibrium for $\epsilon = 0$ and out of equilibrium for $\epsilon > 0$.
A finite $\epsilon$ causes heat dissipation in the system and the production of entropy in the thermostat~\cite{Patra2016} by regulating the gradient $dT/dq = \epsilon/\cosh^2 q$ driving heat transfer.
In this model, the instantaneous heat loss $-d_t Q(t)$ from the system to the thermostat is the local phase space volume contraction rate~\cite{Hoover2016b,Patra2016}:
\begin{align}\label{eq:lambda}
- \frac{d_t Q(t)}{T} = \Lambda(t) = - \zeta(t)\left(\alpha + \beta\frac{3p(t)^2}{T}\right). 
\end{align}
The sign of $\Lambda$ changes when the trajectory crosses the $\zeta = 0$ plane.
Phase space points for which $\zeta$ is positive (negative) has negative (positive) dissipation rate $\Lambda$.
 A negative (positive) dissipation indicates that heat is lost to (gained from) the thermostat.
At long times, however, the average heat dissipation is negative which means that heat is transferred from the system to thermostat and the production of entropy.

\begin{figure}[t!]
	\hspace{-0.5cm}\includegraphics[width=1\columnwidth]{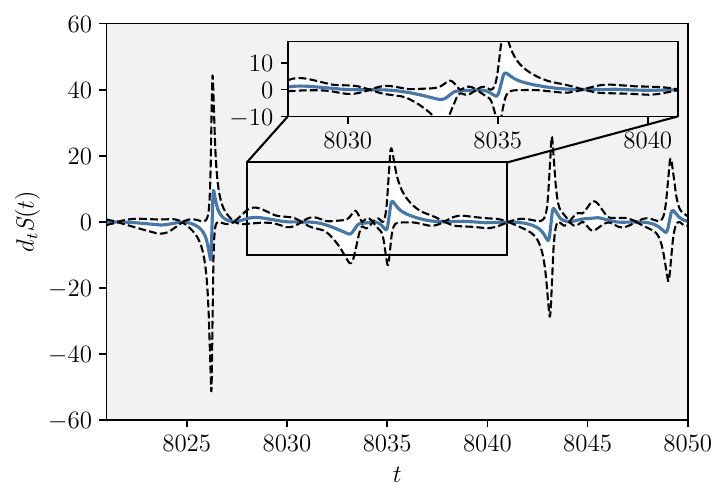}
	\caption{\label{fig:0532-avg-lambda} Time evolution of the entropy rate $d_t\textit{S}(t)$, in the steady state of the 0532 model. The trajectory originates from the $\zeta = 0$ plane. For the bounds, the extremal eigenvalues of the symmetric part of the stability matrix obtained from Eq.~\ref{eq:thermostat_0532} are time averaged. We set the parameter $\epsilon = 0.25$ and $T = 1$. The shaded region cannot be accessed by $d_t\textit{S}$. \textit{Inset} --- Magnified interval $8028 \leq t \leq 8041$ to show the sharpness of the bounds.} 
\end{figure}

For simulations, we set $\epsilon = 0.25$ and $T = 1$, so that the system is operating out of equilibrium.
To compute $\Lambda$ and the bounds, we start a trajectory on the $\zeta = 0$ plane which evolves in the three dimensional phase space $(q, p, \zeta)$. We simulated $10^8$ time steps of size $10^{-3}$ time units for a total time of $10^5$ reduced time units. 
It typically shows transient behavior before relaxing to the steady state in long times.
Figure~\ref{fig:0532-lambda} shows the time evolution of $\Lambda(t)$, which is the instantaneous heat loss $d_t{Q}$, and the bounds during a transient phase of this example trajectory.
The bounds are the extremal eigenvalues of $\stability_+$ computed from Eq.~\ref{eq:thermostat_0532}.
The inset Fig.~\ref{fig:0532-lambda}(a) gives an enlarged view of the evolution and bounds in the time interval $33\leq t\ \leq 49$.
While both the bounds on $\Lambda(t)$ hold along the trajectory, we observe that the lower bound tends to be tighter from above and below when there are smaller deviations from zero. 

The time evolution of $\Lambda$ along a trajectory for the 0532 model usually shows an intermittent behavior with spikes and quiescent phases.
Two of those spikes are visible in  Fig.~\ref{fig:0532-lambda} around $t= 30$ and $t = 50$.
As shown in Fig.~\ref{fig:0532-lambda}(b) these spikes in $\Lambda(t)$ are across the $\zeta = 0$ plane and indicate that the direction of heat exchange, \textit{from} or \textit{to} the system, switches quickly.
At longer times, however, the system relaxes to a steady state and fluctuations in $\Lambda(t)$ are smaller compared to the transient phase fluctuations.
Heat exchange continues to change directions depending on the sign of $\Lambda(t)$.

In the steady state, the phase space trajectories tend to converge to an attractor.
In this steady state, it is sufficient to evolve one trajectory for long times.
To compute the entropy rate and bound on it in this state, we first discard an initial transient part of the trajectory (a time period of $5\times 10^2$).
After the system has relaxed, we calculate the negative of $\Lambda(t)$ (and bounds), which then represents $d_t\textit{S}(t)$ in the steady state of the thermostat at the temperature gradient $\epsilon = 0.25$.
Figure~\ref{fig:0532-avg-lambda} shows the time evolution of $d_t\textit{S}(t)$ bounded by $- \sigma^{\stability_+}_{\text{min}}(t)$ (above) and $-\sigma^{\stability_+}_{\text{max}}(t)$ (below) in the steady state.
We have magnified the evolution in a small time interval to show the sharpness of the bounds, Fig.~\ref{fig:0532-avg-lambda} inset.
As expected, $d_t\textit{S}(t)$ does not a definite sign at a given $t$.
However, the asymptotic time average $\overline{d_t\textit{S}}^\infty$ remains positive as one expects for the entropy production rate.

Because the thermostat is set at $T = 1$, instantaneous, finite and asymptotic time bounds on $-\Lambda$ are also bounds on the heat transfer rate to the thermostat.
The lower bound on $\langle \Lambda\rangle$ sets the maximum heat that can theoretically be dissipated to the surroundings at any time $t$.
Similar bounds on thermodynamic entropy production, and therefore on heat transfer, might also be found in other thermostats that are not necessarily ergodic~\cite{Hoovers2020}.

\subsection{Isokinetically thermostatted Lorentz gas}

As a second example, we consider the Lorentz gas -- a time-reversible system of charged particles moving in a 2D array of fixed hard disk scatterers in the presence of an external electric field $\boldsymbol{E}$.
This system is widely used as model to investigate nonequilibrium steady states and to compute transport coefficients~\cite{dorfman1999introduction}.
This driven Lorenz gas interacts with a thermal reservoir that maintains a fixed kinetic energy of the scattered particles, an isokinetic Gaussian thermostat~\cite{Klages2000}. For this and 
Nos\'e-Hoover thermostats, some transport coefficients, such as electrical conductivity, are simple functions of the sum of Lyapunov exponents (referred to as the ``Lyapunov sum rule''~\cite{Klages2007}).

For any dynamics subject to a Gaussian thermostat like the Lorentz gas, we can use Eq.~\ref{eq:Entropy_prod_prod_space_avg} to put bounds on transport coefficients.
Here, the system is isokinetically thermostatted by applying a frictional force that maintains a constant kinetic energy of the charged particle.
Its equation of motion is
\begin{align}\label{eq:thermo_system}
\dot{\boldsymbol{r}} & = \boldsymbol{p}, \quad \quad \dot{\boldsymbol{p}}= q\boldsymbol{E} - \alpha\boldsymbol{p},
\end{align}
where $\alpha=q\boldsymbol{E}\cdot\boldsymbol{p}/|\boldsymbol{p}|^2$ is a  friction coefficient.
Its ensemble average value $\langle{\alpha}\rangle$ and the electrical conductivity $\eta$ of the system are related to the entropy flow rate:
\begin{align}
  d_t\textit{S} = k_B\langle{\alpha}\rangle = \frac{|\boldsymbol{E}|^2 }{T}\eta.
\end{align}
Here $k_B$ is Boltzmann's constant and $T$ the temperature of the thermal reservoir.
We use Eq.~\ref{eq:Entropy_prod_prod_space_avg} to place bounds on the conductivity at any time $t$:

\begin{align}\label{eq:elec_cond}
-\langle\sigma^{\stability_+}_\text{max}(t)\rangle\leq K\eta(t)\leq -\langle\sigma^{\stability_+}_\text{min}(t)\rangle,
\end{align}
where $K = |\boldsymbol{E}|^2/4NT$ and $N$ is the total number of charged particles in a two dimensional physical space. 
In the asymptotic time limit, we have an upper limit on the average value of the conductivity:
\begin{align}\label{eq:elec_cond-avg}
 K\bar{\eta}^\infty\leq -\overline{\sigma^{\stability_+}_\text{min}}^\infty.
\end{align}
Thus, the bounds on the electrical conductivity of the system are set by the average maximal eigenvalues of $\stability_+$ computed from linearizing the dynamics given by Eq.~\ref{eq:thermo_system}. For a large driving field strength, the system is in a highly nonlinear regime wherein the conductivity $\eta$ is an irregular function of $\boldsymbol{E}$~\cite{Lloyd1995,Klages2000}. In this non-Ohmic regime, a reliable estimation of $\eta$ is a challenge. However, the upper and lower bounds above are still computable, even for an external electric field $\boldsymbol{E}$ that is varying in time, and predict the range of possible values of $\eta$ at a given driving field.

\section{Conclusions}\label{sec:Discussion}

We derived upper and lower bounds on instantaneous Lyapunov exponents and their finite-time counterparts within the classical density matrix theory of dynamical systems. 
These bounds are the extremal eigenvalues of the symmetric part of the stability matrix, which are applicable to deterministic systems evolving continuously in time.
These results are the basis for bounds on the phase space dissipation rate, which hold during both energy accumulation and energy dissipation, as illustrated by the van der Pol oscillator.
Averaged over a statistical ensemble, these bounds on macroscopic quantities derive directly from the properties of the underlying microscopic dynamics. 
For example, the upper bound is the maximum rate of entropy production, heat dissipation, and electrical conductivity. 

\begin{acknowledgments}

This material is based upon work supported by the National Science Foundation under Grant No.\ 2124510. 
S.D.\ thanks Walter Lai for help with code optimization and helpful conversations with Dr. Puneet Patra about the 0532 model. 
The authors acknowledge the use of the supercomputing facilities managed by the Research Computing Group at the University of Massachusetts Boston.

\end{acknowledgments}
\appendix

\bibliography{references}

\end{document}